\documentclass{article}
\usepackage{natbib}
\usepackage{authblk}
\usepackage[a4paper]{geometry}
 \usepackage{breakurl}

\usepackage{amssymb}
\usepackage{amsmath}
\usepackage{amsthm}
\usepackage{xcolor}
\usepackage{hyperref}

\usepackage{listings}
\begin{document}




\title{The information mismatch, and how to fix it\\ 
%
}


\author{Samuel J. Weisenthal}
\author{Amit K. Chowdhry}

\affil[1]{University of Rochester Medical Center, 601 Elmwood Ave, Rochester, NY, 14642
}
\affil[ ]{samuel\_weisenthal@urmc.rochester.edu}
\maketitle
\section*{Abstract}

 We live in unprecedented times in terms of our ability to use evidence to inform medical care. For example, we can perform data-driven post-test probability calculations. However, there is work to do.  As has been previously noted, sensitivity and specificity, which play a key role in post-test probability calculations, are defined as unadjusted for patient covariates.  In light of this, there have been multiple recommendations that sensitivity and specificity be adjusted for covariates.  However, there is less work on the downstream clinical impact of unadjusted sensitivity and specificity. We discuss this here. We argue that unadjusted sensitivity and specificity, when mixed with covariate-dependent pre-test probability scores (which are more easily available nowadays given the multitude of online calculators), can lead to a post-test probability that contains an ``information mismatch.'' We write the equations behind such an information mismatch and discuss the steps that can be taken to fix it.

\section{Introduction}

\label{sec:intro}
We live in unprecedented times in terms of our ability to use evidence to inform medical care.   The ability to use Bayes' rule \citep{bayes1958essay} to calculate a data-driven post-test probability is an example of the progress we have made.  However, we have work to do.
Consider a scenario in which a patient presents with swelling of the left calf. After clinical evaluation, the Wells' \citep{wells1998use} pretest probability of deep vein thrombosis (DVT) is computed, and an ultrasound of the left lower extremity is obtained.  Given the ultrasound result, the sensitivity and specificity of the ultrasound, and the pre-test probability of DVT from the Wells score, a post-test probability of a DVT is computed.  
Unfortunately, although this might sound reasonable, this post-test probability will be incorrect, except under unusual circumstances.  

The scenario we have just described is standard in the prevailing medical diagnostic reasoning framework.  However, the post-test probability has an error, because sensitivity and specificity do not depend on patient information, or covariates, such as age, symptoms, signs, etc. For example, expositions of the prevailing medical diagnostic reasoning framework, which define sensitivity and specificity in such a way, see, e.g., \citet{ledley1959reasoning},\citet{parikh2008understanding}, or \citet{trevethan2017sensitivity}.  Although sensitivity and specificity being unadjusted for patient covariates is often taken as a definition in the literature, it is actually an assumption.  In particular, it is the assumption that test results are independent of all patient characteristics other than disease status (for a discussion, see Section 1.1.3 of \citet{pearl2009causality}). 

That such an assumption is questionable has been noted by multiple authors. \citet{HarrellBBR}, Chapter 19, gives a particularly rich description of the corresponding literature. For example, \citet{dawid1976properties} discusses it in light of selection bias,  \citet{hlatky1984factors} for exercise electrocardiography, and \citet{moons1997limitations} for coronary disease tests as a function of, e.g., patient covariates obtained from history and physical examination. \citet{guggenmoos2000validity} criticizes the classical view that sensitivity and specificity are considered ``universal constants''(i.e., constant with respect to patient covariates). The diagnosis chapter in \citet{harrell2001regression}, along with \citet{moons2003sensitivity}, aligns with this and suggests that the role of sensitivity and specificity should be de-emphasized in diagnostic studies, and that direct modeling of the post-test probability, adjusted for covariate information, should be performed instead.  This direct modeling is undertaken, for example, by \citet{janssens2005new}. The argument for sensitivity and specificity depending on patient covariates is best framed in the language of probability. Therefore, some authors \citep{harrellclinicians, 10.1093/jrsssa/qnad038} have recommended that test characteristics should be described using conditional probability notation, as in \citet{pagano2022principles}, rather than plug-in equations.   

Despite the work listed above, highly cited papers on diagnostic testing still define sensitivity and specificity to not depend on covariates.  This is so ingrained in the literature that even large language models appear to assume it is true \citep{weisenthal2023chatgpt}.  Clearly, even though many have called for a paradigm shift in the way we conceptualize the relationship between sensitivity, specificity, and patient covariates, this shift has not occurred. 

In this work, therefore, we give further motivation for change. In particular, we describe the downstream clinical impact of unadjusted test characteristics.   We show that this can yield a post-test probability that combines a pre-test probability that depends on patient information (such as one from a covariate-dependent pre-test probability calculator) with sensitivities and specificities that do not.  We label this the ``information mismatch.'' 
Although this type of information mismatch may appear benign, and it may be so in certain scenarios, it causes the general mathematical formulation of the post-test probability to break down. In this work, we write the equations that show the breakdown and describe how it might be fixed. 

\section{Diagnostic problem}
\label{sec:diagnostic}
In a standard diagnostic problem, we seek to obtain a post-test probability of a disease given a test result. In the prevailing medical diagnosis framework, which depends on Bayes' rule, if $Dz$ is a binary random variable representing disease presence when set to 1 and disease absence, when set to 0, and $Test$ is a binary random variable representing a positive test result when set to 1 and a negative test result when set to 0, we have that
\begin{align}
\label{eq:standard}
\notag p(Dz=1|Test=1) &= \frac{p(Test=1,Dz=1)}{p(Test=1)}\\
\notag &= \frac{p(Test=1|Dz=1)p(Dz=1)}{p(Test=1)}\\
&=\frac{(\text{sensitivity})(\text{pre-test probability})}{p(Test=1)},
\end{align}
where sensitivity is $p(Test=1|Dz=1)$ and pre-test probability is $p(Dz=1)$. 

Using the law of total probability, the denominator can be further written in terms of sensitivity, as defined above, specificity (which is formally defined as $p(Test=0|Dz=0)$) and pre-test probability:
\begin{align*}
p(Test=1)
&=p(Test=1,Dz=1)+p(Test=1,Dz=0)\\
&=p(Test=1|Dz=1)p(Dz=1)\\
&\ \ \ +p(Test=1|Dz=0)p(Dz=0)\\
&=p(Test=1|Dz=1)p(Dz=1)\\&\ \ \ +(1-p(Test=0|Dz=0))(1-p(Dz=1))\\
&=(\text{sensitivity})(\text{pre-test probability})\\&\ \ \  +(1-\text{specificity})(1-(\text{pre-test probability})).
\end{align*}
In this way, one can write the standard post-test probability as a function of sensitivity, specificity, and pre-test probability. 

Note that the expression in (\ref{eq:standard}) only conditions on the test result, as do most examples from the medical diagnosis literature (see the commonly cited overviews of post-test probability, such as
\cite{parikh2008understanding} or \citet{trevethan2017sensitivity}).  Conditioning only on the test result does not align with clinical practice. The clinician wishes to provide the most patient-specific post-test probability, and this probability is a function not only of the test result but also of the other patient covariates (i.e., other information we have about the patient). 

Unlike  ({\ref{eq:standard}}), consider the case where one also conditions on patient covariates, $X.$ For example, $X$ might contain symptoms or other information, such as age, sex assigned at birth, etc. Conditional on $X=x$, we obtain the post-test probability
\begin{align}
\label{eq:ptpConditionOnx}
\notag p(Dz=1|Test=1,x)&=\frac{p(Test=1|Dz=1,x)p(Dz=1|x)p(x)}{p(Test=1|x)p(x)}\\&\notag =\frac{p(Test=1|Dz=1,x)p(Dz=1|x)}{p(Test=1|x)}
\\&=\frac{(\text{sensitivity}_x) (\text{pre-test probability}_x)}{p(Test=1|x)}
\end{align}  
Note that, as was done for the denominator of (\ref{eq:standard}), the denominator of (\ref{eq:ptpConditionOnx}) can be rewritten
\begin{align*}
p(Test=1|x)
&=p(Test=1|Dz=1,x)p(Dz=1|x)\\&\ \ \ +p(Test=1|Dz=0,x)p(Dz=0|x)\\
&=p(Test=1|Dz=1,x)p(Dz=1|x)\\&\ \ \ +(1-p(Test=0|Dz=0,x))(1-p(Dz=1|x))\\
&=(\text{sensitivity}_x)(\text{pre-test probability}_x)\\&\ \ \  +(1-\text{specificity}_x)(1-(\text{pre-test probability}_x))
\end{align*}
Now, in (\ref{eq:ptpConditionOnx}), $\text{sensitivity}_x$ is defined as $p(Test=1|Dz=1,x)$ instead of $p(Test=1|Dz=1),$ and likewise for $\text{specificity}_x$, $p(Test=0|Dz=0,x)$, and $\text{pre-test probability}_x$, $p(Dz=1|x)$. One might argue that, e.g., $\text{sensitivity}_x$ is not a true ``sensitivity,'' which should be unconditional on covariates, or similarly that a post-test probability that conditions on covariates is not a true post-test probability. This ambiguity might cause problems.

In general, in  (\ref{eq:ptpConditionOnx}),  the presence of the covariate, $x,$ requires (\ref{eq:standard}) be extended to the expression in (\ref{eq:ptpConditionOnx}), which represents the conditional distribution of the test (sometimes known as ``sensitivity'') in a different way. 
If the expression is not extended in this way, possibly due to the ambiguity mentioned above, the post-test probability will be incorrect, as we will now describe.

\section{The information mismatch, an example}
Consider a scenario in which a patient presents with swelling in the left lower extremity. Wells' score for a deep venous thrombosis is computed, which gives $p(DVT+|swelling, cancer,etc \dots).$\footnote{Pre-test probability calculations can be easily performed using \citet{mdcalcWellsCriteria}, which is a particularly useful website, but also may sometimes give access to powerful tools without clear instructions on how to rigorously integrate them into the clinical workflow, which is what we will see here} The literature on ultrasound (US) is consulted to find a sensitivity, $p(US+|DVT+)$ and specificity, $p(US-|DVT-)$. For example, one consults \citep{goodacre2005systematic}.  However, note that in \citet{goodacre2005systematic}, we are given $p(US+|DVT+)$ instead of \[p(US+|DVT+, cancer, swelling,etc \dots).\] In other words, the sensitivity and specificity values are not conditional on patient covariates. 

Using these sensitivity and specificity values in conjunction with Wells' score, as done above, implicitly assumes that
\begin{align*}
&p(DVT+|US+,swelling, cancer,etc...)\\
&=\frac{{\bf p(US+|DVT+)}p(DVT+|swelling, cancer,etc...)}{p(US+|swelling, cancer,etc...)}.
\end{align*}
However, the last expression is usually false. It should be 
\begin{align*}
&p(DVT+|US+,swelling, cancer,etc...)\\
&=\frac{{\bf p(US+|DVT+,swelling, cancer,etc...)}p(DVT+|swelling, cancer,etc...)}{p(US+|swelling, cancer,etc...)}.
\end{align*}
In general, the post-test probability is incorrect when one implicitly assumes
\begin{align*}
p(Dz=1|Test=1,x)=\frac{{\bf p(Test=1|Dz=1)}p(Dz=1|x)}{p(Test=1|x)},
\end{align*}
where the denominator is assumed to be written as
\begin{align*}
p(Test=1|x)&=p(Test=1|Dz=1)p(Dz=1|x)\\ &\ \ \ + p(Test=1|Dz=0)p(Dz=0|x).
\end{align*}
 The last two displays only hold if $Test \perp X,$ in which case   $p(Test=1|Dz=1)$ equals $p(Test=1|Dz=1,x).$    
 
However, it is rarely the case that $Test \perp X.$  It is unlikely that US is independent of the severity of the DVT, where the severity is taken into account in the Wells' score based on the tenderness of the leg and the extent of the swelling, since more severe disease is often easier to detect \citep{harrell2001regression}. That this independence between the test and covariates does not hold for other medical conditions has been shown in many studies; see, for example, \citet{moons1997limitations},\citet{hlatky1984factors},\citet{prince2021evaluation}, or the other examples listed in \citet{moons2003sensitivity},\citet{harrell2001regression}, and \citet{guggenmoos2000validity}. 

The prevailing diagnostic framework, however, draws users into assuming that $Test\perp X$ by naming the special case, $p(Test=1|Dz=1)$, as sensitivity, and then invoking sensitivity when discussing post-test probabilities that depend on covariate-specific pre-test probabilities, such as the Wells' score. 

Note, for example, that in virtually all expositions of the prevailing medical diagnosis framework (see e.g. \citep{parikh2008understanding,trevethan2017sensitivity})
the authors define 
\begin{align}
\label{plugin}
\text{sensitivity}= \frac{\text{true positives}}{\text{true positives + false negatives}},
\end{align}
which is $p(Test=1|Dz=1)$.  
But $p(Test=1|Dz=1)$ is a special case, because sensitivity can be more broadly defined as including $\text{sensitivity}_x$, $p(Test=1|Dz=1,x).$ If sensitivity truly depends on patient covariates, only $p(Test=1|Dz=1,x)$ is useful to obtain the true clinical desirable, $p(Dz=1|Test=1,x).$
 
\section{Clinical impact}

As shown above, the most significant consequence of the information mismatch occurs under a formally calculated post-test probability. Currently, for many problems, this is not done (although we believe it should be). Sometimes, however,  a Fagan nomogram \citep{fagan1975nomogram} is used, which is equivalent to a post-test probability calculation (by transforming probabilities to odds and using a diagram) and is thus subject to the information mismatch. 
 
In current times, it is more common that a post-test probability be informally estimated. Often,  it is noted that a patient has a high pre-test probability of DVT via Wells' score, an ultrasound is ordered, and then it is concluded that the patient has a DVT or not according to the US result in light of the sensitivity and specificity from the literature. This informality may somewhat protect from the consequences of the information mismatch, because there is not too much confidence placed in a numerically calculated post-test probability.  However, it is important to note that the sensitivity and the specificity of the ultrasound from the literature do not take into account patient covariates, and therefore still do not match in an information sense with the Wells' score.  So, even an informal line of reasoning that combines these two measures is subject to the information mismatch.

As stated above, our hope is that calculations for post-test probabilities will become commonplace. In such a case, the information mismatch must be avoided.  As we argue in the next section, in line with other authors, the best way to do so might involve a change in the way we think about diagnosis.

\section{Fixing the information mismatch}

 The easiest way to avoid the information mismatch is to assess whether $Test\perp X.$ If $Test\perp X,$ one can use the unadjusted sensitivity and specificity, as before.  If not, which is expected based on the literature mentioned in Section \ref{sec:diagnostic}, things must be done differently.  For one, new and redesigned studies that estimate sensitivities and specificities that depend on patient covariates could be conducted.  I.e., the estimand of interest would be $\text{sensitivity}_x,$ $p(Test=1|Dz=1,x)$ instead of $p(Test=1|Dz=1).$  Then it would be possible to correctly compute (\ref{eq:ptpConditionOnx}). For example, when performing a study to evaluate the sensitivity and specificity of an ultrasound, one would collect the patient covariates that are present in the Wells' score, and one would report $\text{sensitivity}_x$ and $\text{specificity}_x$ adjusted for those covariates, making it possible to match the information in the pre-test Wells' score with the information in the test characteristics. When performing these redesigned studies, it is important to recognize that terminology is crucial (e.g., one must differentiate sensitivity from $\text{sensitivity}_x).$ 

Some authors argue that we might be able to avoid this conversation about sensitivity altogether.  For example, in \citet{harrell2001regression}, it is recommended that post-test probability, $p(Dz=1|test,x)$, be modeled directly, without the use of Bayes' rule. One would then have a risk score for the pre-test probability, $p(Dz=1|x)$, and an additional risk score for the post-test probability, $p(Dz=1|test,x)$.  Evaluating the impact of a test would be tantamount to considering the change in post-test probability that  occurs when moving from $p(Dz=1|x)$ to $p(Dz=1|test,x).$ Concretely, one would compare Wells' score with an augmented Wells' score that also takes into account ultrasound result. One would therefore be able to avoid getting into the weeds about sensitivity versus $\text{sensitivity}_x.$ This offloads much of the probabilistic responsibility to those who conduct the studies, but it would require a plethora of new studies and easy access to a variety of risk calculators (see the forum comment \citet{Boback} for a discussion on this point). 

While performing new studies to estimate $\text{sensitivity}_x,$ or bypassing sensitivity altogether by modeling the post-test probability directly, help avoid the information mismatch, they do not address the root cause, and they do not help us ensure the best use of existing tools. Due to the increasing prevalence of patient-specific risk scores like Wells', and the ease with which they are now computed, concepts like sensitivity, specificity, and pre-test probability, which were originally meant for screening, have now become integrated into complex clinical scenarios.  However, the language of screening is not sufficient for these clinical scenarios.  The information mismatch might thus be a manifestation of a larger issue with the language we use to describe screening and diagnosis, which includes ambiguous terms like ``sensitivity,'' which may or may not be interpreted as ``$\text{sensitivity}_x.$''

To truly fix the information mismatch,  probabilistic concepts must be described in the language of probability.  As discussed in \citet{harrellclinicians} and \citet{10.1093/jrsssa/qnad038}, it is crucial that there be less emphasis on terminology such as ``sensitivity" and its formula in (\ref{plugin}) --- or, for this matter, ``$\text{sensitivity}_x,$'' which could be, against our recommendations, represented as a new formula --- and more emphasis on the probability behind these concepts, and how to derive and evaluate probabilistic statements.  
Ultimately, this is most likely to fix the information mismatch, and will greatly enhance the application of evidence to clinical problems. 

\section{Discussion}

We have discussed the clinical consequences associated with the assumption that sensitivity and specificity do not depend on patient covariates. Namely, we have shown that it can lead to an information mismatch between pre-test probabilities that depend on patient covariates with sensitivities and specificities that do not.  We have shown the equations related to the information mismatch using a real-world example, and we have discussed the implications for clinical care.  We have also given steps to fix the information mismatch, both in terms of redesigning studies and in terms of changing the lens through which diagnostic concepts are viewed.

We build on a rich lineage of work describing the issues with sensitivity and specificity in medical diagnosis. This literature has been driven forward in large part, especially recently, by Harrell, who, although not directly involved in the current work, influenced its framing greatly through his papers and online writing.  Our work is more of a synthesis of the existing literature than a novel insight, but we do, unlike existing work, begin with the clinical consequence and work backward. Existing work criticizes sensitivity and specificity as probabilistic concepts, ostensibly taking it for granted that the downstream consequences will be obvious. However, this is only the case if one has a strong command of probability, the time to connect the dots, and the motivation to do so. We facilitate these connections, and we motivate attention to the problem by showing the clinical impact directly from the outset, and even naming it---the information mismatch (drawing a connection to the well-known problem of ventilation/flow (V/Q) mismatch). In alignment with the problem-focused nature of medicine, we identify a problem, which has been exacerbated by the more-readily available pre-test probability calculators.  

Beyond identifying a problem, we present a solution, which is also more of a synthesis of existing proposals than a novelty in its own right. However, we diverge slightly from critics of sensitivity and specificity, who might prefer that a solution focus less on $sensitivity_x$ and more on the posterior probability of disease, $p(dz+|test+,x).$ However, we maintain that $sensitivity_x$ would be a step toward solving the information mismatch, and even more, its discussion facilitates the development of insight into the probabilistic nature of test characteristics, and---as we believe that critics of sensitivity and specificity would agree---this insight is perhaps more important than anything else. 

Note also that the information mismatch might  occur if sensitivity and specificity  are surreptitiously adjusted for covariates, such as when a study estimates ``sensitivity’’ in a subset of the population.  For example, many studies restrict their cohorts to adults or patients with severe disease.  This could lead to a particularly difficult-to-detect information mismatch. Ultimately, regardless of the etiology, the information mismatch will only become more common with the increase in online pre-test probability calculators,  and the medical diagnosis framework must be updated to fix this.

\section{Acknowledgments}
The authors appreciate discussions on \url{https://discourse.datamethods.org/}, in particular, on the topics \textit{Sensitivity, specificity, and ROC curves are not needed for good medical decision making} (\url{https://shorturl.at/HwiKz}) and \textit{Determining post-test probability of Covid-19} (\url{https://shorturl.at/UoIdR}).
This research, which is the sole responsibility of the authors and not the National Institutes of Health (NIH), was supported by the National Institute of General Medical Sciences (NIGMS) under T32GM007356 and T32GM152318.

\bibliographystyle{elsarticle-harv} 
\bibliography{bib}

\end{document}